\def\pr#1 {Phys. Rev. {\bf D#1\tie }}
\def\pe#1 {Phys. Rev. {\bf #1\tie}}
\def\pre#1 {Phys. Rep. {\bf #1\tie}}
\def\pl#1 {Phys. Lett. {\bf #1B\tie }}
\def\prl#1 {Phys. Rev. Lett. {\bf #1\tie }}
\def\np#1 {Nucl. Phys. {\bf B#1\tie }}
\def\ap#1 {Ann. Phys. (NY) {\bf #1\tie }}
\def\cmp#1 {Commun. Math. Phys. {\bf #1\tie }}
\def\imp#1 {Int. Jour. Mod. Phys. {\bf A#1\tie }}
\def\mpl#1 {Mod. Phys. Lett. {\bf A#1\tie}}
\def\zp#1 {Z. Phys. {\bf C#1\tie}}
\def\amp#1 {Adv. Theor. Math. Phys. {\bf#1\tie}}
\def\jhep#1 {JHEP {\bf #1\tie\rm }}
\def\tie{\noexpand~}
\def\ov{\overline}
\def\s{(\sigma)}

\def\be{\begin{equation}}
\def\ee{\end{equation}}
\def\bea{\begin{eqnarray}}
\def\eea{\end{eqnarray}}
\def\Ref#1{(\ref{#1})}

\catcode`@=11
\def\marginnote#1{}
\newcount\hour
\newcount\minute
\newtoks\amorpm
\hour=\time\divide\hour by60
\minute=\time{\multiply\hour by60 \global\advance\minute by-\hour}
\edef\standardtime{{\ifnum\hour<12 \global\amorpm={am}%
        \else\global\amorpm={pm}\advance\hour by-12 \fi
        \ifnum\hour=0 \hour=12 \fi
        \number\hour:\ifnum\minute<10 0\fi\number\minute\the\amorpm}}
\edef\militarytime{\number\hour:\ifnum\minute<10 0\fi\number\minute}
\def\draftlabel#1{{\@bsphack\if@filesw {\let\thepage\relax
   \xdef\@gtempa{\write\@auxout{\string
      \newlabel{#1}{{\@currentlabel}{\thepage}}}}}\@gtempa
   \if@nobreak \ifvmode\nobreak\fi\fi\fi\@esphack}
        \gdef\@eqnlabel{#1}}
\def\@eqnlabel{}
\def\@vacuum{}
\def\draftmarginnote#1{\marginpar{\raggedright\scriptsize\tt#1}}
\def\draft{\oddsidemargin 0.0truein
        \def\@oddfoot{\sl preliminary draft \hfil
        \rm\thepage\hfil\sl\today\quad\militarytime}
        \let\@evenfoot\@oddfoot \overfullrule 3pt
        \let\label=\draftlabel
        \let\marginnote=\draftmarginnote
   \def\@eqnnum{(\theequation)\rlap{\kern\marginparsep\tt\@eqnlabel}%
\global\let\@eqnlabel\@vacuum}  }
\catcode`@=12
\documentstyle[epsf, 12pt]{article}
\begin{document}

\thispagestyle{empty}

\bigskip\bigskip
\begin{center}
\Large{\bf SYMMETRY BREAKING IN STRING THEORY}
\end{center}

\vskip 1.0truecm

\centerline{\bf Ioannis  
Giannakis\footnote{giannak@summit.rockefeller.edu}}
\vskip5mm
\centerline{\it Physics Department}
\centerline{\it Rockefeller University}
\centerline{\it 1230 York Avenue}
\centerline{\it New York, NY 10021}

\vskip5mm

\bigskip \nopagebreak \begin{abstract}
\noindent
In this paper we discuss symmetry breaking in string theory.
Spacetime symmetries are implemented as inner automorphisms of
the underlying superconformal algebra. Conserved currents generate
unbroken spacetime symmetries. As we deform the classical solutions
of the string equations of motion, the deformed currents continue
to generate spontaneously broken symmetries eventhough they cease to
commute with the string Hamiltonian. We illustrate these ideas
by studying supersymmetry breaking in a non-trivial string background.
\end{abstract}

One of the outstanding problems of string theory is
to understand the symmetry principles underlying the theory.
This is a non-trivial problem because, unlike most physical
theories, string theory has been formulated from the
beginning as a set of rules for calculating scattering
amplitudes. The problem is akin to discovering the full (spontaneously
broken) $SU(2) \times U(1)$ gauge invariance of electroweak interactions
from the Feynmann rules of the Standard Model.

There are persuasive reasons to believe that string theory
does possess a much richer symmetry sructure and it is surely
important to understand it. The interactions are unique, presumably
fixed by symmetry, scattering amplitudes exhibit universal behaviour
\cite{gross} and the discovery of
an infinite symmetry algebra for the bosonic
string theory \cite{mgn} strongly suggest the existence
of an enormous underlying symmetry.

A general formalism for determining the spacetime symmetries
of string theory was presented in \cite{meov}. This formalism
relates deformations of the BRST operator to spacetime symmetries.

To each solution of the string equations of motion corresponds
a superconformally invariant two-dimensional field theory.
This theory will be completely defined by specifying its
BRST operator as a local function of world-sheet fields and their
canonical momenta. The spacetime fields appear as the couplings
of this two-dimensional field theory. Thus,
in string theory, the BRST operator $Q_{\Phi}$ is parameterized
by spacetime fields $\Phi$. 

If there is a transformation of the spacetime fields that is
a symmetry of string theory, then to every solution of the
equations of motion there will be a new solution, where the fields
take their transformed values. Thus to every superconformal field theory
there will correspond another, with transformed couplings.
Furthermore, since they are related by a symmetry, these two
solutions should be physically indistinguishable. Two superconformal
field theories are isomorphic if they are isomorphic as
operator algebras, that is if there is a map
\bea
\rho: {\cal A}_1 \mapsto {\cal A}_2 
\eea
between the operator algebras ${\cal A}_1$ and ${\cal A}_2$
of the two superconformal field theories that maps the BRST operator
of the one theory to the BRST operator of the other and
preserves the equal time commutation relations.
For any algebra we may construct another algebra isomorphic to
the first one, by means of an infinitesimal similarity
transformation, also called inner automorphism. That is, take
${\cal A}_1={\cal A}_2$ and
\bea
\rho_h(a)=a+i[h, a]
\eea
where  h is any fixed, infinitesimal operator.

For any infinitesimal operator $h$, then, the superconformal
field theories specified by $Q_\Phi$ and $Q_{\Phi+\delta\Phi}$
are isomorphic. Thus if
\bea
i[h, Q_\Phi]=Q_{\Phi+\delta\Phi}-Q_\Phi={\delta}Q
\eea
for some $\delta\Phi$, it follows that $\Phi \mapsto
\Phi+\delta\Phi$ is a symmetry of the spacetime fields.

We may clarify the way in which the change in the BRST operator
may be interpreted as a change in the spacetime fields by first
considering the more general problem of deforming a superconformal
field theory. How does the SCFT deform as we deform the classical
solution of the string equations of motion 
\cite{nontrivial,ovr, mg}? In general this
deformation will not correspond to a symmetry transformation, it is
a physically different, nearby solution. For example, in
General Relativity, two nearby solutions are flat Minkowski space
and a weak gravitational wave propagating through it. We seek then
deformations of the BRST operator $Q \mapsto Q+\delta Q$
that preserve nilpotency
\bea
\{ Q, \delta Q \}=0
\eea
It is straightforward to show that to first order nilpotency is
preserved by deforming the BRST charge with
\bea
\delta Q=\int d{\sigma}cV(\sigma)
\label{eqav}
\eea
where $V(\sigma)$ is a BRST invariant operator.

The second question is ``What subset of the above deformations
do correspond to symmetry transformations?''. If we take the
generator $h$ to be the zero mode of an 
infinitesimal $(\frac{1}{2}, 0)$ and
$(0, \frac{1}{2})$ superprimary field, then it
is straightforward to see that its
action on the BRST charge  is necessarily of the form of Eq.
\Ref{eqav}
and thus can be interpreted as a change in the spacetime fields.

It is well known that conserved currents \cite{dixon, gm}
generate symmetries, but within the formalism described here,
conservation is not necessary, a fact that does not seem to
have been widely appreciated. Indeed it is not hard to
demonstrate that a non-conserved current generates a spacetime
symmetry that is spontaneously broken by the particular
background.

In the remaining of this paper we shall discuss how
supersymmetry is implemented as an inner automorphism
of the operator algebra. Initially we shall consider string propagation
on $M^7 \times T^3$, where $M^7$ is 7-dimensional Minkowski
spacetime and $T^3$ a 3-dimensional torus. Spacetime supersymmetry
is generated by a conserved two-dimensional current. As we
deform our classical solution by turning on
a constant $H=dB$ flux along the toroidal dimensions-shifting
the vacuum-spacetime supersymmetry is spontaneously broken
and is implemented by a non-conserved two dimensional current.

The operator that generates 7-dimensional supersymmetry transformations
about the background $M^7 \times T^3$
in the heterotic string is \cite{she}
\bea
h={\int}d{\sigma}{\epsilon}^{{\alpha}a}(X)
S_{{\alpha}a}e^{-{\frac{\phi}{2}}}\s=
{\int}d{\sigma}{\epsilon}^{{\alpha}a}(X)J_{{\alpha}a}\s
\label{eqperez}
\eea
where $S^{{\alpha}a}$, $e^{-{{\phi}\over 2}}$,
are the spin fields
for the two-dimensional fermions $\psi_\mu(\sigma)$, and
the superconformal ghosts $\beta(\sigma), \gamma(\sigma)$
respectively. We have written the operators in a 7-dimensional
notation, $\alpha=1, \cdots 8$ are the $SO(7)$ spinor indices
and $a=1, 2$ an internal index. We shall also use the following
decomposition of the Dirac matrices
\bea
\Gamma^{\mu}=\frac{i}{\sqrt 2}
({\gamma}^{\mu}\otimes 1 \otimes {\sigma^1}), \qquad
\Gamma^{i}=\frac{i}{\sqrt 2}(
1\otimes {\sigma^i} \otimes {\sigma^2})
\eea
where $\gamma^\mu$ satisfy the Clifford algebra
in $7$ dimensions $\{ \gamma^\mu, \gamma^\nu \}=2\eta^{\mu\nu}$
and $\sigma^i$ are the Pauli matrices.

The currents $J_{{\alpha}a}\s$ are conserved $[H, J_{{\alpha}a}\s]=0$,
where $H$ is the two-dimensional string Hamiltonian.
The integrand is superprimary of dimension $(\frac{1}{2}, 0)$ only if the
parameters of transformation $\epsilon^{{\alpha}a}(X)$
satisfy
\bea
{\gamma}^{\mu}{\partial_{\mu}}{\epsilon^{{\alpha}a}}=0.
\eea
Let's calculate then the commutator of $Q$ with
$h$ given by Eq. \Ref{eqperez}. We find
\bea
\delta Q=i[h, Q]={\int}d{\sigma}c{\partial_\lambda}{\epsilon}^{{\alpha}a}
S_{{\alpha}a}e^{-{{\phi}\over 2}}{\ov\partial}X^{\lambda}.
\label{eqriot}
\eea
A spacetime symmetry is unbroken when the vacuum values of the fields
are invariant under the transformation. In our formalism this
would imply that $\delta Q=0$ and this corresponds to the condition
that a spacetime symmetry is unbroken by the
particular string background. What would then this imply
for supersymmetry in $M^7$? We observe
that this condition is satisfied only if the parameter of 
supersymmetry transformations $\epsilon^\alpha$ is constant.
This confirms then the well known result that global $N=2$ supersymmetry
is unbroken in the string background $M^7 \times T^3$.

Next we deform our classical solution by turning a constant
$H_{ijk}$ flux along the three compact dimensions of $T^3$.
Thus our background consists of $M^7 \times T^3$ and a nonzero
flux. 
This nontrivial string background is described in terms of a
nilpotent BRST operator $\hat Q=Q+{\delta}Q$, where $Q$ is the
BRST operator that describes string
propagation on $M^7 \times T^3$ and $\delta Q$ the nilpotent
deformation that describes the nontrivial
$H_{ijk}$ flux \cite{bag},
\bea
\delta Q={\int}d{\sigma}({1\over 2}cB_{ij}
{\partial}X^{i}{\ov\partial}X^{j}
+{1\over 2}c{\partial_k}
B_{ij}{\psi}^{k}{\psi}^{i}{\ov\partial}X^{j}
-{1\over 4}{\gamma}B_{ij}{\psi}^{i}{\ov\partial}X^{j})\s.
\eea

The generator of supersymmetry transformations in the particular
background becomes
\bea
{\hat h}={\int}d{\sigma}{\epsilon}^{{\alpha}a}(
S_{{\alpha}a}e^{-{{\phi}\over 2}}+{1\over
4}{\sigma}^{ij}B_{ij}
S_{{\alpha}a}e^{-{{\phi}\over 2}})\s={\int}d{\sigma}
\epsilon^{{\alpha}a}{\hat J}_{{\alpha}a}\s.
\label{eqcit}
\eea
The deformed supersymmetry currents ${\hat J}_{{\alpha}a}$
cease to be conserved with respect to the deformed Hamiltonian,
$[{\hat H}, {\hat J}_{{\alpha}a}] \ne 0$ but continue
to generate supersymmetry transformations about
the particular background. The integrand remains
a superprimary field of dimension $(\frac{1}{2}, 0)$ if
the parameters of transformation obey the following conditions
\bea
{\gamma}^{\rho}{\partial_\rho}{\epsilon}^{{\alpha}a}
+{m\over 2}{\epsilon}^{{\alpha}a}=0,
\eea
where $H_{ijk}=m\epsilon_{ijk}$.

Furthermore we can calculate the commutator of $\hat Q$ with
$\hat h$, Eq. \Ref{eqcit}. The result is
\bea
\delta Q=i[{\hat h}, {\hat Q}] ={\partial_{\lambda}}{\epsilon}^{{\alpha}a}
S_{{\alpha}a}e^{-{\frac{\phi}{2}}}
{\ov\partial}X^{\lambda} & - & \frac{im}{2}{\sigma_i}{\epsilon}^{{\alpha}a}
S_{{\alpha}a}e^{-{\frac{\phi}{2}}}
{\ov\partial}X^{i} \\ \nonumber
& + & \frac{1}{4}{\partial_{\lambda}}
{\epsilon}^{{\alpha}a}{\sigma}^{ij}B_{ij}
S_{{\alpha}a}e^{-{\frac{\phi}{2}}}{\ov\partial}X^{\lambda}
\eea
We observe that it is imposible to have $\delta Q=0$ for any
value of the parameters of transformation $\epsilon^\alpha$.
This indicates that $N=2$ supersymmetry is spontaneously broken
by the non-zero vacuum value of the $H$ flux.
Furthermore this formalism can be used to describe in detail both
the Higgs and the SuperHiggs mechanisms in string theory
\cite{bg}, \cite{superhiggs}.

\section{Acknowledgments}

This paper is based on work I have done in collaboration
with J. Bagger.
This work was supported in part by the Department of Energy Contract
Number DE-FG02-91ER40651-TASKB.


\begin{thebibliography}{99}
\bibitem{gross}{D. Gross and P. Mende, \np{303} \ 407 (1988).}
\bibitem{mgn}{M. Evans, I. Giannakis and D.
Nanopoulos \pr{50} \ 4022 (1994).}
\bibitem{meov}{M. Evans and B. Ovrut,
\pr{41} \ 3149 (1990)}
\bibitem{nontrivial}
{J. Freericks and M. Halpern, \ap{188} \ 258 (1988).}
\bibitem{ovr} { B. Ovrut and S. Kalyana
Rama, \pr{45} \ 550  (1992);\\
 J. C. Lee, \zp{54} \  283  (1992).}
\bibitem{mg}{M. Evans and I. Giannakis, \pr{44} \ 2467 (1991).}
\bibitem{dixon}{T. Banks and L. Dixon, \np{307} \ 93 (1988).}
\bibitem{gm}{M. Evans and I. Giannakis, \np{472} \  139  (1996);\\
I. Giannakis, \pl{388} \  543 (1996).}
\bibitem{she}{D. Friedan, E. Martinec and S. Shenker, \np{271} \ 93 (1986).}
\bibitem{bag}{J. Bagger and I. Giannakis, \pr{65} \ 046002 (2002).}
\bibitem{bg}{J. Bagger and I. Giannakis, \pr{56} \ 2317 (1997)}
\bibitem{superhiggs}{J. Bagger and I. Giannakis, hep-th/0502107}


\end{thebibliography}
\end{document}